\documentclass[12pt]{article}

\usepackage{amssymb,amsmath,latexsym,graphics,authblk,url,setspace,caption}

\usepackage{graphicx}
\usepackage{epstopdf} 
\usepackage{enumerate}
\usepackage{float}
\usepackage{color}
\usepackage{amssymb}
\usepackage{verbatim}
\usepackage{setspace}
\usepackage{amsmath}
\usepackage{hyperref}
\usepackage{algorithm}
\usepackage{algpseudocode}
\usepackage{color}
\usepackage{footnote}
\usepackage{threeparttable}
\usepackage{caption}
\usepackage{setspace}

\usepackage[square,sort,comma,numbers]{natbib}

\usepackage[portrait, margin=0.8in]{geometry}

\title{Practical privacy metrics for synthetic data}
\author{Gillian M Raab, Beata Nowok \& Chris Dibben}

\renewcommand{\baselinestretch}{1.5} 

\begin{document}

\maketitle
\begin{abstract}
This paper explains how synthpop package \footnote{see \url{www.synthpop.org.uk}} has been extended to include functions to calculate measures of identity and attribute disclosure risk for synthetic data that measure risks for the 
records used to create the synthetic data. The basic function \texttt{disclosure} calculates identity disclosure for a set of quasi-identifiers (keys) and attribute disclosure for one variable specified as a target from the same set of keys. The second function \texttt{multi.disclosure} is a wrapper for the first and presents summary results for a set of targets. This short paper explains the measures of disclosure risk and documents how they are calculated. We recommend two measures: $RepU$ (replicated uniques) for identity
disclosure and $DiSCO$ (Disclosive in Synthetic Correct Original) for attribute disclosure. Both
are expressed a \% of the number of original records and each can be compared to similar measures calculated from the original data on its own.
Experience with using the functions on real data found that some apparent disclosures could be identified as coming from relationships 
in the data that would be expected to be known to anyone familiar with its features. We flag cases when this seems to have occurred and provide means of excluding them.
This paper was written as a vignette for the R package synthpop version 1.9-3.
\end{abstract}
\section{Introduction}
\renewcommand{\baselinestretch}{1.5} 
In his recent review of synthetic data (SD) methodology Reiter \cite{reiter2023}
comments: 
\begin{quote}
''While there is need to examine disclosure risks in synthetic data, there is no standard for
doing so, especially in fully synthetic data. Instead, disclosure checks tend to be ad hoc"
\end{quote}
This is in contrast to the variety of measures of utility available for SD; see \cite{raab2021} for a list of these. While utility measures must be chosen that are relevant to the intended use of the SD, disclosure measures must focus on the possible harm to the privacy of an individual or other unit whose data contributed to the creation of the SD. The evaluation of the disclosure risk from SD must relate to the context of its release (see \cite{elliotanonframe} for a discussion of this). We cannot expect a fixed rule, for example that a criterion for release requires a value of some disclosure measure beyond a threshold. Instead, we expect those releasing data to use the disclosure measures to evaluate potential harm to the data subjects, or to the data custodians, from information in the released data. We hope that these new functions will allow the disclosure risk of SD sets to be explored and, where necessary, reduced. 

We expect that disclosure risks from SD to be lower than those from the original data. But in some cases, e.g. a large data set with only a few categorical variables, the disclosure risk from the original may already be low. To evaluate the protection from disclosure afforded by a synthesis method, the risks for SD must be compared with equivalent risks for the original ground truth data (GT). We consider two types of disclosure risk:

\begin{itemize}
\item{\textit{identity disclosure}: This refers to the ability to identify individuals in the data from a set of known characteristics that we will refer to as keys. Identity disclosure may be less relevant for completely synthesized\footnote{Complete or full synthesis is when all values of all variables are replaced by synthetic values. This is in contrast to incomplete or partial synthesis where only some variables are replaced.} data because there is no one-to-one correspondence between records in the GT and the SD. But it may still be of interest since it is an important factor for attribute disclosure.}
\item{\textit{attribute disclosure}:This refers to the ability to find out from the keys something, not previously known, for an attribute associated with a record in the GT.}
\end{itemize}

The measures we describe here are appropriate for fully SD where all items in all records are replaced by synthetic values. Different measures have been developed for partially SD \cite{DrechslerReiter2009,ReiterMitra2009}. Note that our disclosure methods treat numeric variables, by default, as if they were categories unless \texttt{ngroups\_keys} and/or \texttt{ngroups\_target(s)} are set. In our first example the \texttt{income} variable has been grouped into 20 categories, but the other numerical variable \texttt(depress) with only 21 categories has been left as it is.  The number of groups formed may differ from the parameter setting e.g. if there are fewer than \texttt{numgroups} distinct values.\footnote{The code uses grouping options from functions in the \texttt{classInt} package.}. 

The disclosure risk posed by SD can be reduced by using techniques from statistical disclosure control (\textbf{sdc}), such as aggregation of categories, smoothing of numeric values or removal of replicated uniques. These methods can be used to reduce disclosure risk by modifying the GT before synthesis, or the SD before its is released. Some such methods are already available in the \textbf{synthpop} package. These include categorising, top/bottom coding and smoothing for continuous variables, and the merging of small categories for factors. The removal of replicated uniques is another option available in \texttt{synthpop}.

There have been many recent proposals for making synthetic data sets comply with  Differential Privacy (DP) \cite{Dwork2006b}. DP is a very strong privacy guarantee that protects against an intruder with arbitrary external information about the subjects in the data, except for the one whose privacy is being protected. This is an unrealistic assumption and DP SD has been shown to have unacceptably low utility in many cases \cite{bowen_ss, groundhog}. We will not discuss these methods here, but note that we could use the metrics proposed here to evaluate disclosure risks for DP SD. 

The Appendix introduces the notation we will use to define the disclosure measures that can be computed by the disclosure functions in the $synthpop$ package. Our approach involves creating a composite variable, $q$, from the  quasi-identifiers (keys) for the original and synthetic data.  Identity disclosure measures are calculated from univariate tables of $q$ for the GT and the SD. Attribute disclosure measures for a target variable $t$ are obtained by tabulating $t$ against $q$ for the GT and the SD.  For our initial example in Section \ref{sec:simpexamp} and most of the later development we assume thet the 
number of records created in the SD ($N_s$) is the same as the number in the GT ($N_d$). The exception is Section 3.3.2 where we show how different proposed measures for attribute disclosure are affected by reducing the number of synthetic records ($N_s$) that are released.

\section[A simple example]{A simple example}\label{sec:simpexamp}
Here we illustrate the basic use of \texttt{multi.disclosure}. If the parameter \texttt{targets} is not specified, all the variables in the SD that are not part of keys are used as targets.
The identity disclosure measures are \texttt{UiO} for the original and \texttt{repU} for synthetic, and for attribute disclosure \texttt{Dorig} for original and \texttt{DiSCO} for synthetic. These and other measures will be explained in Sections \ref{subsec:ident} and \ref{subsec:attrib}.

\begin{figure}[ht]
  \centering
  \includegraphics[width = 1\linewidth]{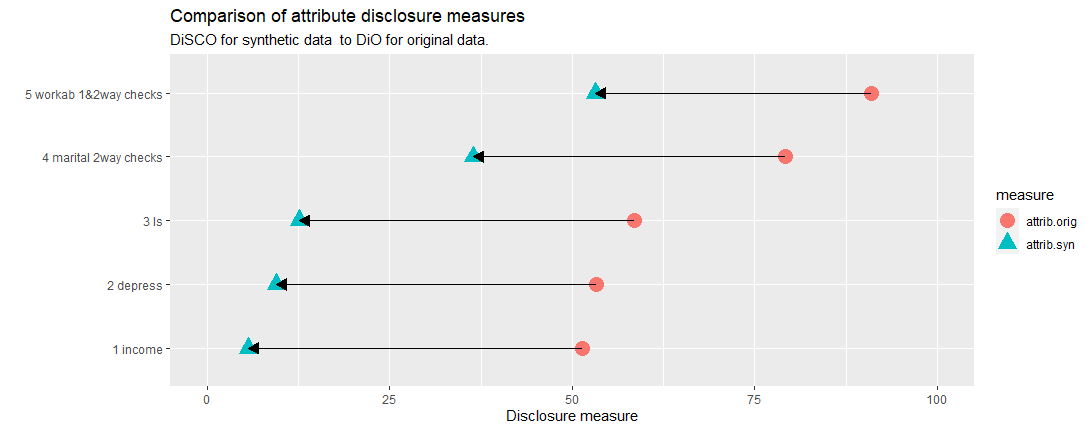}
  \caption{Plot from \texttt{multi.disclosure} with \texttt{plot = TRUE}, the default value.}
  \label{fig:f1}
\end{figure}

First, a subset of 9 variables are selected from the \texttt{SD2011} data (a survey on quality of life in Poland) that is available as part of the \texttt{synthpop} package. A single synthetic data set is created by the default method in \texttt{synthpop}: \texttt{cart} for each conditional distribution. Note that the variable \texttt{income} has values -8 that indicate not applicable, and the synthesis allows for this. The synthetic data object \texttt{s1}\footnote{an object of class \texttt{synds}} has a component \texttt{syn} that is a single synthetic data set. The disclosure functions can also be used with synthetic data created by other methods either as single synthetic data sets or lists of repeated syntheses from the same original. Here we select 4 keys that represent items that might be known
for members of this sample, or of the Polish population in 2011. By default,the 5 other variables become the targets: \texttt{depress} \texttt{income} \texttt{ls} \texttt{marital} (marital status), \texttt{workab} (intention to work abroad). The second target, \texttt{income}, is grouped with a target of 20 categories\footnote{The algorithm attempts to create this number of groups with approximately equal sizes. Skew data often result in fewer groups. In this case 18 groups are formed from the numeric values, plus two groups from the -8 code and the missing values, giving a total of 20}
We first use \texttt{multi.disclosure} to create a \texttt{multi.disclosure} object for these keys, and we print out its identity component.
\renewcommand{\baselinestretch}{1.0}
\begin{verbatim}
R> library("synthpop")
R> ods <- SD2011[, c("sex", "age", "region","placesize","depress",
+    "income","ls","marital" , "workab")]
R> s1 <- syn(ods, seed = 8564, print.flag = FALSE, cont.na = list(income = -8))
R> t1 <- multi.disclosure(s1, ods, print.flag = FALSE, plot = FALSE, 
+  keys = c("sex", "age", "region", "placesize"),   ngroups_targets = c(0,20,0,0,0))
R> print(t1,to.print = "ident")

Disclosure risk for 5000 records in the original data

Identity disclosure measures
from keys: sex age region placesize 
For original  ( UiO )  48.38 %
For synthetic ( repU ) 14.86 %.

\end{verbatim}
\renewcommand{\baselinestretch}{1.5}
The measure $UiO$ (Unique in Original) shows that 48\% of the original records would have unique combinations of these 4 keys. The term, ''singling out" is used in data protection regulation for this type of attribute disclosure\footnote{See for example its use in the UK Information Commissioner's guidance on anonymisation here https://ico.org.uk/media/about-the-ico/documents/4018606/chapter-2-anonymisation-draft.pdf}. For the synthetic data $RepU$ tells us that almost 15\% of the original records would be unique in the original and also unique in the synthetic data.
For attribute disclosure we can examine the results as either a table or a plot. Here we see in Figure \ref{fig:f1} the plot that would have been generated if \texttt{plot} had been set to TRUE in the code above, the attribute disclosure results would have been plotted as shown in Figure \ref{fig:f1}. Details of how identity and disclosure measures are calculated can be found in Section 3. The results in the Figure 1 are in descending order by the attribute disclosure risk in the SD.

The measure $Dorig$ tells us that these 4 keys would identify a unique value of each of these targets for over 50\% of the original records. We get lower values for $DiSCO$, the proportion of the original records that are disclosive in the original and also in the synthetic data with a correct attribution to the target. The first two variables shown in Figure \ref{fig:f1} have additional labels that 
flag possible contributions to disclosure from knowledge of 1-way or 2-way relationships in the original data. This is detailed in Section 4.

\section{Scenario and definitions}\label{sec:sc_def_not}
\subsection{Setting the scene}\label{subsec:scen}
These disclosure measures are intended to assess what a person who only has access 
to the synthetic data can infer about known individuals who are present in the original data. We will use the term "intruder" for such a person, though no malicious intent is implied. The intruder is assumed to have information for one or more individuals about the value of certain key variables that are present in the same format in the original and the synthetic data. They attempt first to see if the individual is present, and then to determine the value of other items in the data file that we refer to as targets. We are assuming a worst-case scenario where the intruder believes 
they are querying the original data.\footnote{This may not be too unrealistic if the data are made available inadvertently, or if the intruder thinks that efforts to label the synthetic data as e.g. "Fake Data" are thought just to be a cover up. It may also be a reqasonable measure to use without this scenario, since it compares the disclosiveness of the SD to that of the GT.}
Disclosure measures from the synthetic data are each compared to similar measures for someone with access to the original data. Here we will introduce the measures by an example. Formal definitions with notation and formulae are in the Appendix.
The first step in evaluating disclosure risk, as described here, is to identify a set of keys that might be expected to be known to an intruder. These keys are then combined to form a quasi-identifier that we designate as $q$. For example, if we have hospital records we might define age,sex date and hospital as keys and this would give a $q$ with levels such as ''\texttt{78 | M | 1/1/2024 | WG}" for a 78 year old man admitted to hospital WG on 1/1/2024.

\subsection{Identity disclosure measures}\label{subsec:ident} 
The concept of k-anonymity is central to identity disclosure for microdata. First proposed in 1998 \cite{kanon1} it is discussed fully in \cite{elliotanonframe}. A table is k-anonymous if a set of keys identifies at most $k$ individuals. Thus 2-anonymous data will never identify just one individual. Based on this idea, the percentage of records for which the keys identify just one individual give identity disclosure measures. 
 Tables of $q$ values are produced from the SD and the GT. $UiO$ and $UiS$ are the percentages of records with keys where the table count is 1. An intruder checking out a record for their known set of keys will look for it in the synthetic data. Some records
will not be in the SD and $UiOiS$ (Unique in Original in Synthetic) gives the \% that would be found. These records are then checked for uniqueness in the
synthetic data giving, $repU$ is the percentage of unique records in GT that are also unique in the SD.

The percentage $repU$ has been used as a disclosure measure to evaluate SD by \cite{jackson_rss} and by \cite{raab22} \footnote{Jackson et al. in \cite{jackson_rss} argue that the denominator for $repU$ should be $N_s$ rather than $N_d$. This is inappropriate because our scenario is to consider the risk to the original data, as we discuss for attribute disclosure in Section 3.2.2.}.
Replicated uniques are used in \texttt{synthpop} as part of the statistical disclosure control function, \texttt{sdc}, that includes the option of reducing disclosure risk by removing them from the SD. Nowok et al. \cite{nowok_repu} have evaluated this and give an example where this process has very little effect on utility. The function \texttt{replicated.uniques}\footnote{For example by \texttt{replicated.uniques(s2, ods, keys = c("sex","region","age","placesize"))}} also calculates $repU$ using a different method from the one described here.

One of the outputs of the functions \texttt{disclosure} and \texttt{multi.disclosure} is \texttt{ident}, a table of identity disclosure measures as illustrated by this example, using the keys from the example in Section \ref{sec:simpexamp} but now calculated for a synthetic object with 5 data sets, using the one target \texttt{depress}. We first create \texttt{t5} an object of class \texttt{disclosure} and then print out the identity and attribute disclosure measures for each synthetic data set.

The table for identity disclosure has $UiO$ and $repU$ as its first and last column. The 2nd and 3rd columns are $UiS$ calculated from the synthetic data in the same manner as $UiO$ from the original and $UiOiS$ the \% of $UiO$ with $q$ that are in the SD but not necessarily unique. Each are steps towards calculating $repU$.
\renewcommand{\baselinestretch}{1.0}
\begin{verbatim}

R> s5 <- syn(ods, seed = 8564, m = 5, print.flag = FALSE)
R> t5 <- disclosure( s5, ods, keys = c("sex", "age", "region",
+    "placesize"), target = "depress", print.flag = FALSE)
R> print(t5, to.print = c("ident"))


Disclosure measures from synthesis for 5000 records in original data.

Identity disclosure measures for 5 synthetic data set(s) from keys:
 sex age region placesize 

    UiO   UiS UiOiS  repU
1 48.38 37.34 22.68 14.86
2 48.38 35.44 22.24 13.96
3 48.38 35.18 21.98 13.62
4 48.38 34.90 22.08 13.78
5 48.38 36.14 22.00 14.62

\end{verbatim}
\renewcommand{\baselinestretch}{1.5}

\subsection{Attribute disclosure measures}\label{subsec:attrib} 

The attribute disclosure measures  displayed in Figure 1 are our preferred measures, $DiSCO$, for the SD and $Dorig$ for the GT \cite{raabPSD2024}. In section 3.3.1 we explain them via the example above. Other related disclosure measures, $DCAP$ and $TCAP$ and their variants, have been proposed .  We prefer $DiSCO$ over $TCAP$ and its variants \cite{little2022,little2025} because it measures all the ways synthesis is effective in reducing attribute disclosure.  A further measure  $DCAP_d$, a modification of
\cite{elliot2014SYLLS,taub_PSD2018}, also measures all these aspects.  The measures $DiSCO$ and $DCAP_d$ are  included in the $attrib$ component of the output from the disclosure function. For completeness, all the other measures are computed by $disclosure$ and returned in the $allCAPs$ component of the disclosure object.

The two measures $DiSCO$ and $DCAP_d$ are the only two measures that show how the disclosure risk is reduced by releasing SD with fer records than GT, see Figure2. They can be considered as the extremities of a range of a more general disclosure measure when
 a record is considered disclosive if its probability of correct attribution reaches a threshold $tau$. $DiSCO$ corresponds to a $tau$ of 1.0 while $DCAP_d$ corresponds to a $tau$ of zero. This more general approach may be included in future releases of $synthpop$. 
 Section 3.3.2 discusses all these other measures and formal definitions of all measures are in the Appendix.

\subsubsection{Disclosive in Synthetic Correct in Original: $DiSCO$}

We approach disclosure risk from the point of view of an intruder with access to the SD
and to certain attributes (quasi-identifiers) known for one or more individuals in the original data.
The quasi-identifiers for the known record(s) can be combined to create a composite variable $q$ that can be created for all records in the original and SD.
Modelling what an intruder might do we calculate the following measures, each of which is a proportion of the original records:
\begin{itemize}
\item{Look up $q$  in the SD for each original record in the GT. The proportion found becomes $iS$ (in Synthetic).}
\item{Check if all records with the same $q$ have the same level of the target $t$ in the SD. The proportion of all GT records passing this further test becomes $DiS$ (Disclosive in Synthetic).}
\item{Then check if these apparent disclosures correspond to the value of $t$ in the original data. The proportion of original records for which this is true
becomes $DiSCO$ Disclosive in Synthetic Correct Original.}
\end{itemize}
Note that records contributing to $DiSCO$ may not be disclosive in the original data, as this information would not be available to the intruder. A further measure $DiSDiO$ (Disclosive in Synthetic Disclosive in Original) restricts the score to those disclosive in the original. 
\renewcommand{\baselinestretch}{1.0}
\begin{verbatim}

R> print(t5, to.print = c("attrib"))

Disclosure measures from synthesis for 5000 records in original data.

Attribute disclosure measures for depress from keys: sex age region placesize 
  Dorig  Dsyn    iS   DiS DiSCO DCAPorig DiSDiO max_denom mean_denom
1  53.3 46.26 64.90 34.18  9.54    16.39   6.14         3       1.16
2  53.3 44.80 64.00 32.50 10.26    17.45   6.78         4       1.19
3  53.3 44.60 64.02 32.14  9.10    16.20   5.92         4       1.19
4  53.3 45.80 63.88 33.38  9.20    15.92   5.52         4       1.21
5  53.3 44.52 63.44 31.46  9.34    16.17   5.80         4       1.23

\end{verbatim}

\renewcommand{\baselinestretch}{1.5}
Here we print the table of attribute disclosure measures for the example 
in the previous section.
As we move from $iS$ to $DiS$ and to $DiSCO$  we can see how the \%disclosive is affected by different conditions. Here lack of $q$ levels in the SD retains just 64\% of records. The requirement for a record to be disclosive in
the SD reduces this to 34\% and again to around 9\% for those with a correct attribution. This reduces it to around 6\% by restricting to 
records disclosive in the original.

The $Dorig$ and $DiSCO$ measures are not restricted to disclosures that are identified from unique records for $q$ in either the original or the SD. The number of records contributing to each disclosive $qt$ cell in the synthetic table is the denominator that applies to that record. The columns \texttt{max denom} and \texttt{mean denom} refer to the denominators for the records disclosive in the original data that contribute to $DiSCO$. We can see from the mean that here the majority of disclosive records had unique key combinations in the SD, and the maxima was 3 for the first synthesis and 4 for the others. Large denominators can be an indication that some of the disclosure may be coming from strong relationships between variables in the data that might even be expected 
a-priori. This aspect is discussed further in Section \ref{sec:onetwoway}. The disclosure measures can be restricted to those with small denominators by using
the parameter \texttt{exclude\_over\_denom\_lim} to TRUE. The example in Section \ref{sec:simpexamp} is here run restricted to denominators of 1.
\renewcommand{\baselinestretch}{1.0}
\begin{verbatim}

R> multi.disclosure(s1, ods, print.flag = FALSE, plot = FALSE,
+    keys = c("sex", "age", "region", "placesize"),
+    denom_lim =1, exclude_ov_denom_lim = TRUE)


Disclosure risk for 5000 records in the original data

Identity disclosure measures
from keys: sex age region placesize 
For original  ( UiO )  50.26 %
For synthetic ( repU ) 16.36 %.

Table of attribute disclosure measures for sex age region placesize 
Original measure is  Dorig and synthetic measure is DiSCO 
Variables Ordered by synthetic disclosure measure

                     attrib.orig attrib.syn                   check1 Npairs
1 income                   48.38       3.36                               0
2 depress                  48.38       7.02                               0
3 ls                       48.38       9.84                               0
4 marital                  48.38      15.40                               0
5 workab 1way checks       48.38      20.48 Check  workab  level  NO      0
                     check2
1 income                   
2 depress                  
3 ls                       
4 marital                  
5 workab 1way checks       

\end{verbatim}
\renewcommand{\baselinestretch}{1.5}

The $DiSCO$ values have decreased, as expected, but $UiO$ and $repU$
have increased and also that $Dorig$ is now the same for all targets and equal to the $UiO$ value before the denominator exclusion. This makes sense because 
removing large denominators from $UiO$ and $repU$ increases the number of uniques. Also, with denominators of 1, all unique values of $q$ are disclosive in the original data.

Note that decreases in $DiSCO$ are more pronounced
for \texttt{workab} and \texttt{marital}, the two targets that were flagged to check 1-way or 
2-way relationships. Their original $DiSCO$ values were 37\% for \texttt{marital} and 53\%
for \texttt{workab}. This and other methods of excluding certain apparent disclosures 
are discussed in the next section.

\subsubsection{Other attribute disclosure measures}\label{subsubsec:other} 
Previous work on attribute disclosure \cite{elliot2014SYLLS,taub_PSD2018} has used the Correct Attribution Probability ($CAP$) as a disclosure measure for synthetic data. This is the proportion for a record in the SD where its $q$ correctly predicts the attribute level of the target, $t$, in the original. Two versions are described, each with the total $CAP$ in the numerator, but different denominators. Some combinations of the keys, $q$, found in the SD may not exist in the original, so that their $CAP$ is undefined. One version, that we will designate as $DCAP_{b}$, ignores the records with undefined probabilities and uses as a denominator $N_{b}$, the number of synthetic records where the value of $q$ occurs in the original.\footnote{This approach is sometimes referred to as $DCAP_{undefined}$} \cite{taub_PSD2018}.  A second version ($DCAP_{s}$) sets the undefined attribution probabilities to zero for records where $CAP$ is undefined and uses as denominator  ($N_s$) the total number of synthetic records. Note that $DCAP_{b} \ge DCAP_s$ because $N_{b} \le N_s$.  If the total mumber of original records is used as the dsenominator we get the measure $DCAP_d$. This is the same as $DCAP_s$ when the SD has the same number of records as the GT, but will be lower when smaller SD samples are generated.

The related quantity $TCAP$, and its variants. restricts the probabilities to those over a certain limit, and most commonly this limit is $1.0$. With this choice the numerator of the $TCAP$ measures identical to that for $DiSCO$, the proportion of synthetic records correctly  predicted with certainty from the SD. 
The same two choices of denominator as were used for $DCAP$ could be used. One choice $TCAP_{s}$ would be to use the denominator, $N_{s}$ and the other $TCAP_{b}$ would be to use the denominator, $N_{b}$. Note that when $N_s = N_d$, the measure $TCAP_s$ will be identical to $DiSCO$.  However the $TCAP$ that is defined in \cite{little2022, little2025} does not use either of these denominators, but uses as a denominator $N_{bp}$ the number of GT records with $q$ present in the SD and where any level of the target $t$ is predicted with certainty. These are the records identified at the second bullet point of the procedure described in Section 3.3.1 above. The numerators and denominators used for each of these measures are summarised in Table 1.

\begin{table}[h]
\centering
\begin{tabular}{|c|c|c|}
  \hline
   {}  & \multicolumn{2}{|c|}{Numerator} \\
  \hline
    & Sum of proportions correct  & Total number predicted with certainty \\
Denominator & CAP  & CAP = 1\\
\hline
 $N_d$ & $DCAP_d$ & $DiSCO$ \\
  $N_s$ & $DCAP_s$ & $TCAP_s$ \\
  $N_{b}$ & $DCAP_{b}$ & $TCAP_{b}$ \\
  $N_{bp}$ &  & $TCAP$ \\
\hline
\end{tabular}
\caption{Summary of numerators and denominators for $DCAP$ and $TCAP$ measures.}
\label{table:1}
\end{table}

All of these measures are shown in Table 2  for 3 of the targets in the synthesis of the 5,000 records from the example above. We can see that the inequalities 
above all hold. For  this case the ordering of the disclosive risk by target is consistent for all measures.  Because the number of records in the GT and the SD are the same, $DCAP_s$ is identical to $DCAP_d$ and $TCAP_s$ is identical to $DiSCO$. For measures that share the same numerator, Table 1, we see that the reduced denominators have a large effect in this example, so that $TCAP$ is around three times $DiSCO$.

\begin{table}[h]
\centering
\begin{tabular}{rrrrrr}
  \hline
 & $DCAP_s$ and $DCAP_d$ & $DCAP_{b}$ & $TCAP_{d}$ and $DiSCO$ & $TCAP_{b}$ & $TCAP$ \\ 
  \hline
ls & 21.97 & 29.57 & 12.63 & 17.00 & 36.71 \\ 
  depress & 16.23 & 21.85 & 9.21 & 12.40 & 28.57 \\ 
  income & 11.34 & 15.27 & 6.16 & 8.29 & 19.96 \\  
\hline
\end{tabular}
\caption{DCAP and TCAP variants (\%) for 3 targets from the example above  when number of synthetic records equals the number of original records (5000). Average results for 5 SDs.}
\label{table:2}
\end{table}

For all the examples we have examined so far SD has been generated with sample size $N_s$ the same as the number of original records $N_d$; $N_s = N_d$.
Another way that the disclosure risk to the original records can be reduced is to release  synthetic data sets that are smaller than the original. This is the approach that has been used when UK Census data has been released  \cite{DaleSARS} or for low-fidelity Census data from Scotland\footnote{\url{https://www.researchdata.scot/news-and-insights/synthetic-census-data-now-available-for-research/} accessed 7/5/25}.

\begin{figure}[h]
\centering
\includegraphics[width=1\linewidth]{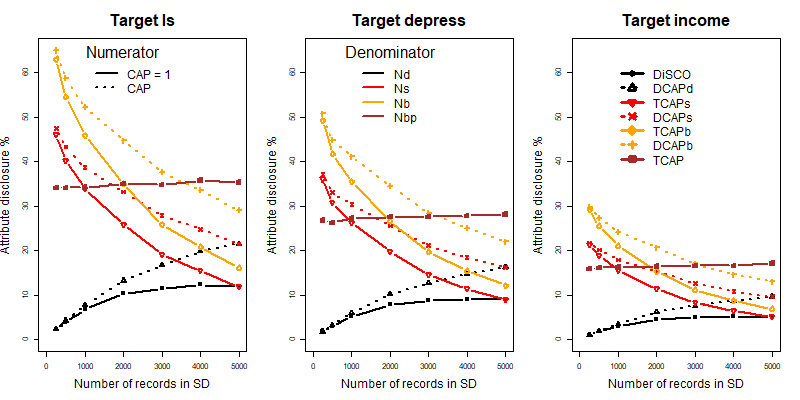}
\caption{Plot of disclosure measures for 3 targets for synthetic data of size ($k$) from 250 to 5000. Note that the legends apply to all figures, with type of line indicating the numerators and colours the denominators of each measure.}
\label{fig:f2}
\end{figure}

Figure 2 illustrates how the different disclosure measures assess the risks in this example for sizes of SD from 5000, as the original, to only 250 records. Of the measures in Table 1, only $DiSCO$ and $DCAP_d$ 
adjust the attribute disclosure risk correctly for the reduced number of records in the SD. 
The $TCAP$ measure is almost independent of the size of the SD. The other measures all give misleadingly high disclosure risks for small samples.We can see how the pairs of 
measures ($DiSCO$, $TCAP_s$) and ($DCAP_d$,$DCAP_s$) diverge for $N_s$ values lower than $N_d$. 

Referring back to section 3.3.1 we can see how
these measures relate to the three bullet points giving the stages by which an intruder takes to identify an attribute disclosure. The measure $TCAP$ assesses the intruder's success rate only at the final bullet point, starting with records in $DiS$. $TCAP_s$ measures the success of the last two steps, from $iS$, but only $DiSCO$ includes the disclosure risk from all 3 steps. 

 \section{Identifying disclosure from 1-way and 2-way relationships}\label{sec:onetwoway}
As mentioned in the Introduction, what we can learn about disclosiveness of attributes can depend on our prior knowledge of the data set or the population from which it is drawn. It would not be practical to specify our prior probability for every possible combination of keys. However, checking two aspects
of disclosure results can help us to check when we might have predicted the correct attribution with high probability without knowledge of all the keys.
The first is a check for a target where a high proportion of records have one level of the target. The second is when there is a strong relationship
between a target and one of the keys, so that one $tq$ pair accounts for many of the disclosive records. These two aspects are flagged by the
values of \texttt{check\_1way} and \texttt{check\_2way} that are returned as part of objects of 
Returning to the example in Section \ref{sec:simpexamp}, we now use the function \texttt{disclosure} 
to get details for the two targets that were flagged as requiring checking; see Figure \ref{fig:f1}. 
\renewcommand{\baselinestretch}{1.0}
\begin{verbatim}

R> d1_workab <- disclosure(s1, ods, print.flag = FALSE, target = "workab",
+     keys = c("sex", "age", "region", "placesize"),plot = FALSE)
R> print(d1_workab, to.print = c("check_1way"))


Disclosure measures from synthesis for 5000 records in original data.

Details of target level contributing disproportionately to disclosure
   Level  All PctLevelAll totalDisclosive nLevelDis PctLevelDis
NO    NO 5000       88.64            2605      2482       95.28

Details of target-key pairs contributing disproportionately to disclosure of workab 
26 pairs need checks 

Please examine component $check_2way of the disclosure object and look at original data.
Consider excluding these key-target pairs with some the following parameters to disclosure:
exclude_ov_denom_lim = TRUE or defining key-target combinations from exclude.targetlevs,
exclude.keys and exclude.keylevs

\end{verbatim}
\renewcommand{\baselinestretch}{1.5}
We can see that it was the category "NO" of \texttt{workab} that contributed most to the
disclosure risk; most survey respondents (89\%) had never worked abroad. This level of
the target accounted for 95\% of apparent disclosures for \texttt{workab}. To
predict this level for all of a group with the same $q$ could hardly be considered 
disclosive. This would be true if the intruder had access to the marginal distribution
of \texttt{workab}, and even if they did not, some knowledge of the respondents' background
might suffice. The $tq$ pairs identified for \texttt{workab} both included the "NO"
level of the target.
\renewcommand{\baselinestretch}{1.0}
\begin{verbatim}

R> d1_marital <- disclosure(s1, ods, print.flag = FALSE, target = "marital",
+     keys = c("sex", "age", "region", "placesize"),plot = FALSE)
R> print(d1_marital, to.print = c("check_2way"))


Disclosure measures from synthesis for 5000 records in original data.

Details of target-key combinations contributing disproportionately to disclosure
This is a list with one for each of 4 6 syntheses
  target_key_levs npairs key key_target_total key_total PctTargetKeyLevel
7       SINGLE|19     10 age               91        92             98.91
4      MARRIED|41      5 age               59        73             80.82
6       SINGLE|18      5 age               91        92             98.91
8       SINGLE|22      5 age               85        91             93.41

\end{verbatim}
\renewcommand{\baselinestretch}{1.5}
For the target \texttt{marital} two $tq$ pairs are identified as contributing large denominators. Those aged 19 are almost all SINGLE, and those aged 40 are mainly MARRIED. Again these would not 
be considered disclosive for an intruder with some knowledge of the respondents.

The thresholds for identifying one-way and two-way relationships can be modified. In each case there are two criteria that need to be satisfied, one number and one \%. For one-way disclosure 
the parameter \texttt{thresh\_1way} has the default value of $c(50, 90)$, meaning that there must be at least 50 disclosive records for one level of a target and that $q$ values including this target must account for over 90\% of all disclosive records. For two way relationships the \texttt{thresh\_2way} has default value $c(5,80)$ and the algorithm first identifies all 
disclosive $tq$ combinations with denominators over 4. It then identifies the level of the key in $q$ with that best predicts this level of $t$ and checks if over 80\% of the disclosive records with this key would have the correct prediction. 

\section{Excluding records}\label{sec:exclusions}
Having identified records where the apparent disclosure is something that would generally be known, one option is to exclude these key-target combinations explicitly from the measures. The following parameters 
for \texttt{multi.disclosure} and \texttt{disclosure} can be used for this.
\begin{itemize}
\item{\texttt{not.target}: All records with levels of the target given by the parameter \texttt{not.target} are excluded from all disclosure measures.}
\item{\texttt{usekeysNA}:This is set to TRUE by default so that missing values are included in all tables. It can be set to FALSE for some or all keys to exclude NAs.}
\item{\texttt{usetargetsNA (multi.disclosure) or usetargetNA (disclosure)}: Similar to the above for target(s) - note can be a vector for multi.disclosure.}
\item{\texttt{exclude.keys, exclude.keylevs and exclude.targetlev}: Three vectors of length the number of key-target pairs to be excluded from all tables.\footnote{For \texttt{multi.disclosure} these parameters are supplied as lists with elements for each target in \texttt{targets}}}
\end{itemize}
\begin{figure}[ht]
\centering
\includegraphics[width=1\linewidth]{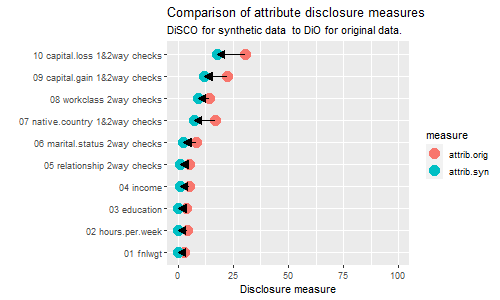}
\caption{Plot from \texttt{multi.disclosure} for variables flagged by checks.}
\label{fig:f3}
\end{figure}
To illustrate exclusions we will use the Adult data set from the UCI machine learning repository \cite{UCI} with almost 50 thousand records from the US Census income study, available as part of the \texttt{arules} package for \textbf{R} \cite{arules}. This is one of the data sets used in \cite{giomi2022anon} to evaluate privacy risks for SD. Note that the two variables, \texttt{capital gain} and textt{capital gain}, originally numeric data, have been truncated so that they have only 123 and 99 distinct values repsectively. They have not been grouped further in the results presented here.  Figure \ref{fig:f3} is the output of \texttt{multi.disclosure} from the keys \texttt{age, occupation, race and sex} for the other 10 variables that are not in the keys.
The disclosiveness of the original data is relatievly low, compared to the previous example, because of the large sample size and the absence of any geographic identifiers. 
Three of the 10 variables \texttt{capital.gain, capital.loss and native.country} are flagged to check one-way relationships. The check1 column of the attribute table tells us that the levels contributing to disclosure are
zeros for the first two and \texttt{United-States} for \texttt{native.country}. These levels make up 95\%, 92\% and 89\% of all records.

Three other variables are flagged as having disclosive two\_way relationships \texttt{workclass, marital, relationship} with totals of 7,7 and 2 $tq$ combinations respectively. The summary function \texttt{multi.disclosure} does not allow target and key specific pairs to be excluded. To
investigate this it is necessary to examine the output of \texttt{disclosure}. This showed that 
the largest contribution to \texttt{check\_2way} for \texttt{workclass} was due to this being missing when occupation was missing, although other
relationships between these two variables also contributed.

\begin{table}[ht]
\centering
\begin{tabular}{|r|rr|rr|rr|rr|rr|}
 \hline
   & \multicolumn{2}{|c|} {} & \multicolumn{2}{|c|} {} & \multicolumn{2}{|c|} {} & 
   \multicolumn{2}{|c|} {denom\_lim 1} &  \multicolumn{2}{|c|} {} \\
  & \multicolumn{2}{|c|} {} & \multicolumn{2}{|c|} {} & \multicolumn{2}{|c|} {NAs out} & 
   \multicolumn{2}{|c|} {NAs out} &  \multicolumn{2}{|c|} {} \\
 & \multicolumn{2}{|c|} {No excl} & \multicolumn{2}{|c|} {not.tlev} & \multicolumn{2}{|c|} {not.tlev} & 
   \multicolumn{2}{|c|} {not.tlev} &  \multicolumn{2}{|c|} {denom\_lim 1} \\
 target & orig & syn & orig & syn  & orig & syn & orig & syn & orig & syn   \\ 
 \hline
capital.gain & 22.55 &  19.87 &  0.21 & 0.00 &  0.21 &  0.00 &  0.19 &  0.00 &  2.68 &  1.17\\
capital.loss &  30.61 &   27.71 &  0.08 &  0.00 &  0.08 &  0.00 &  0.08 &  0.00 &  2.68 &  1.26\\
education &    3.71 & 0.93 &  3.71 &  0.93 &  3.71 &  0.93 &  2.45 &  0.46 &  2.68 &  0.50\\
fnlwgt &  2.70 & 0.00 &  2.70 &  0.00 &  2.70 &  0.00 &  2.45 &  0.00 &  2.68 &  0.00\\
hours.per.week &  4.36 & 0.19 &  4.36 &  0.19 &  4.36 &  0.19 &  2.45 &  0.12 &  2.68 &  0.13\\
income &  4.97 & 2.10 &  4.97 &  2.10 &  4.97 &  2.10 &  1.58 &  0.50 &  2.68 &  0.82\\
marital.status & 8.23 & 5.27 &  8.23 &  5.27 &  8.23 &  5.27 &  2.45 &  0.68 &  2.68 &  0.74\\
native.country & 17.09 &   13.55 &  0.94 &  0.08 &  0.94 &  0.08 &  0.67 &  0.04 &  2.68 &  0.87\\
relationship &  5.17 & 2.64 &  5.17 &  2.64 &  5.17 &  2.64 &  2.45 &  0.69 &  2.68 &  0.73\\
workclass &   14.27 &   11.64 &  14.27 & 11.64 &  14.27 & 11.64 &  2.45 &  0.85 &  2.68 &  0.96\\
  \hline
\end{tabular}
\caption{Disclosure results from Adult data with different exclusions.}
\label{table:3}
\end{table}

Table 2 gives the results of excluding different entries in the tables of $q$ and $t$ from the attribute disclosure measures.
Excluding the levels of the target flagged by \texttt{check\_1way} reduces the $DiSCO$ to almost zero for these 3 variables.
Adding exclusion of missing values reduced the disclosure for \texttt{workclass} and for some other variables a little. 
Adding the restriction to denominators of 1, reduced the disclosure for variables identified by \texttt{check\_2way} to low 
levels. In the final columns we can see that restricting to denominators of 1, by itself, gives low levels of disclosure for
all variables, with the exception of those flagged by \texttt{check\_1way}. This approach was one that was trialed in an earlier 
version of these functions but discarded for giving misleading results for some data. It does not count some attribute disclosures that might well be found by an intruder. For example, if a small number of records with the same keys all had the same level of the target that
corresponded to the level in the original data. We feel that a better approach is to exclude specific target/key combinations.

\section{Conclusions}
The privacy metrics we propose here are in some sense the opposite of differential privacy (DP).
DP claims to protect data from an intruder with arbitrary knowledge of the data, except of
the one record that has the greatest influence on the likelihood of the results. In contrast our metrics require the user to specify keys that identify variables in the data
that would expect to be known about individuals, as well as to 
specify the details of what they would expect an intruder 
to know about the data.  Also, the routines flag cases where part of the
disclosure measures come from one- or two-way relationships so that disclosure would be
expected even in the absence of data. 

We have evaluated the routines on a few data sets
and set levels of the thresholds for this at what seem to be reasonable levels, but 
more experience with other data sets would be valuable. 
We hope that these tools will be helpful to data holders who need to make decisions 
about the risks of releasing SD to the public, or to a restricted audience.
They should also enable the disclosure risks of different synthesis methods to be evaluated.

\section{Acknowledgement}
Research Data Scotland (\url{https://www.researchdata.scot/}) funded Gillian Raab's time to carry 
out the research reported here and  to expand the capabilities of the synthpop package\footnote{
from version 1.8-0 available on CRAN at \url{https://CRAN.R-project.org/package=synthpop}
}
to include measures of disclosure control. We also thank the Scottish Centre for Administrative 
Data Research for continue to support the development of the synthpop package since its creation was
supported by the ESRC funded SYLLS poject in 2012-14.
We would also like to thank Timon Huijser, for pointing out errros in a previous version of this vignette.

\bibliographystyle{acm}
\bibliography{disclosure}

\section*{Appendix}
\subsection*{Notation and formal definitions}{\label{subsec:notation}}
Before defining the measures of identity and disclosure risk we need to introduce the notation that will be used to calculate them. The first step is to create the quasi-identifiers from the keys for the original and SD. For the keys used in the example given in Section \ref{sec:simpexamp} the quasi-identifier that we will designate as $q$ for the first record in the original data is:

\noindent{\texttt{"FEMALE | 57 | Lubuskie | URBAN 100,000-200,000"}}

\noindent{and that for the first record in the SD:}

\noindent{\texttt{"FEMALE | 39 | Zachodnio-pomorskie | URBAN 100,000-200,000"}}.

In order to calculate identity disclosure measures, we need to compare the tables of $q$ from the original and SD. For attribute disclosure measures we need to cross-tabulate $q$ with each target variable $t$ and compare findings from the SD with what would have been found from the original data. In general, the levels of $q$ and sometimes $t$ in the original and SD will not be the same. Before creating any tables, we need to define sets of $q$ and $t$ values that give the union of both sets of levels and align the tables so that their indices correspond.

For the original data $d_{.q}$ is the count of records with the keys corresponding to the levels of $q$ and $q_{tq}$ is the count of records with this $q$ and level $t=1,...T$ of the target. The equivalent counts from the synthesised data are designated by $s_{.q}$ and $s_{tq}$. When a member of $q$ is in the original data but not in the synthetic, $s_{.q}$ and $s_{tq}$ are all zero. Similarly when a member of $q$ is in the SD but not in the original, $d_{.q}$ and $d_{tq}$ are all zero. The two tables can be written as shown in Table 1, where the total records in the original data is $N_d$, made up of $N_{d~only}$ and $N_{d~both}$. The
equivalent totals for the SD are $N_s$, $N_{s~only}$ and $N_{b}$.

\begin{center}
\begin{table}[ht]
\begin{tabular}{ c|ccc|ccc|ccc|c} 
 & \multicolumn{3}{|c|}{only in original} & \multicolumn{3}{|c|}{in both} & \multicolumn{3}{|c|}{only in synthetic} & Total\\
 \hline
1  & ... & $d_{1q}$ & ...  & ... & $d_{1q}$ & ...  & ... & 0 & ... & $d_{1.}$ \\
 ... & ... & ... & ...  & ... & ... & ...  & ... & ... & ... & ... \\
t  & ... & $d_{tq}$  & ...  & ... & $d_{tq}$ & ...  & ... & 0 & ... & $d_{t.}$ \\
... & ... & ... & ...  & ... & ... & ...  & ... & ,,, & ... & ... \\
T  & ... & $d_{Tq}$ & ...  & ... & $d_{Tq}$ & ...  & ... & 0 & ... & $d_{T.}$ \\
 \hline
Column sums  & & $d_{.q}$ &  & ... & $d_{.q}$ & ...  & ... & 0 & ... & $N_d$ \\
 \hline
Totals  &  & $N_{d\:only}$ &  & & $N_{b}$ &  & & 0 &  & $N_d$ \\
\end{tabular}
\end{table}
\begin{table}[ht]
\begin{tabular}{ c|ccc|ccc|ccc|c} 
 & \multicolumn{3}{|c|}{only in original} & \multicolumn{3}{|c|}{in both} & \multicolumn{3}{|c|}{only in synthetic} & Total\\
 \hline
1  & ... & 0 & ...  & ... & $s_1q$ & ...  & ... & $s_1q$ & ... & $s_{1.}$ \\
 ... & ... & ... & ...  & ... & ... & ...  & ... & ... & ... & ... \\
t  & ... & 0 & ...  & ... & $s_tq$ & ...  & ... & $s_tq$ & ... & $s_{t.}$ \\
... & ... & ... & ...  & ... & ... & ...  & ... & ... & ... & ... \\
T  & ... & 0 & ...  & ... & $s_Tq$ & ...  & ... & $s_Tq$ & ... & $s_{T.}$ \\
 \hline
Column sums  & ... & 0 & ...  & ... & $s_{.q}$ & ...  & ... & $s_{.q}$ & ... & $N_s$ \\
 \hline
Totals  &  & 0 &  & & $N_{b}$ &  &  &  & $N_{s\:only}$ & $N_s$ \\

\end{tabular}
\caption{Notation for tables from quasi-identifier ($q$) and target ($t$) from original (upper table) and SD (lower table).}
\label{table:4}
\end{table}
\end{center}
 To calculate the \% of records in the original and SD we need:
  \begin{equation}
\%~Unique~in~Original = UiO = 100\sum{(d_{.q} |d_{.q} = 1)}/N_d. 
  \end{equation}
  \begin{equation}
  \%~Unique~in~Synthetic = UiS = 100\sum{(s_{.q} |d_{.q} = 1)}/N_d.
    \end{equation}
    The intruder has information about the keys for an individual in the real data that they attempt to identify in the SD. They first attempt to find them in the SD, and  the \% found is:
      \begin{equation}
    \%~Unique~in~Original~in~Synthetic = UiOiS = 100 \sum{(d_{.q} = 1 |s_{.q} = 1 \land d_{.q} > 0)}/N_d.
      \end{equation}
      Some of these records would not be unique in the SD, restricting to such records gives:
        \begin{equation}
      \%~replicated~Uniques = repU = 100\sum{(s_{.q} |d_{.q} = 1 \land s_{.q} = 1)}/N_d.
      \end{equation}
      
      To find an attribute from a set of keys, it is necessary to examine the distribution of $s_{tq}$ for groups defined by $q$. We define column proportions for the original and SD as $pd_{tq} = d_{tq}/d_{.q}$ and for the synthetic as $ps_{tq} = s_{tq}/s_{.q}$, and noting that we set $ps_{tq}$ and $pd_{tq}$ to zero whenever their denominators, $d_{.q}$ and $s_{.q}$ are zero.
     
Returning to the scenario described in Section \ref{subsec:scen}, we must first define a measure of attribute disclosure for the original data. This is based on the concept of \textit{l-diversity} \cite{ldiv} that requires that each set of records defined by $q$ has at least $l(\ge{2}$ distinct values of the target. A data set is \textit{l2-diverse} for $q$ and $t$ if all records for every $q$ have the same level of $t$\footnote{This could be generalised to $l>2$, but in practice the levels of targets are not generally exchangeable and a more practical approach would be to aggregate levels for certain targets.} An attribute disclosure measure for the original data can be defined as \textit{\% Disclosive in Original} :
\begin{equation}
 Dorig = 100\sum^q{\sum^t{(d_{tq} |pd_{tq} = 1)}}/N_d.
\end{equation}
The equivalent measure for the SD, taken as if it were real, becomes :
\begin{equation}
 Dsyn = 100\sum^q{\sum^t{(s_{tq} |ps_{tq} = 1)}}/N_s.
\end{equation}
An intruder with access only to the SD, but with knowledge of $q$ from one or more individuals in the original, would look them up in the SD. Some of their $q$ levels be key combinations that do not appear in the SD
leaving the proportion that do appear as $iSO$ (in in Synthetic Original) 
\begin{equation}
 iSO = 100\sum^q{\sum^t{(d_{tq} | s_{tq}>0)}}/N_d.
\end{equation}
A level of $q$ from an original record may identify more than one target in the SD, or identify the wrong target.
To exclude these we require the records to be Disclosive in  the SD and Correct when checked 
with the original giving:
\begin{equation}
 DiSCO = 100\sum^q{\sum^t{(d_{tq} | ps_{tq} = 1)}}/N_d.
\end{equation}

Note that $DiSCO$ can include records that are not disclosive in the original data giving a further 
measure Disclosive in Synthetic and Disclosive in the  Original :
\begin{equation}
DiSDiO = 100\sum^q{\sum^t{(d_{tq} | ps_{tq} = 1 \land pd_{tq} = 1)}}/N_d.
  \end{equation}
As we comment above the intruder would not be able to tell if records were 
identified as $DiSDiO$ rather than $DiSCO$, so we prefer the latter measure.
However, the intruder can identify when the apparently disclosive record
is not unique in the SD. 
 This restriction can be imposed by requiring that the denominator for disclosive records in the original does not exceed a 1,  as described
 and discussed in Section \ref{sec:exclusions}.

\subsection*{DCAP and TCAP measures }{\label{subsec:DCAPTCAP}}

The measures $baseCAPd$, $DCAP$, $TCAP$ are calculated by the function \texttt{disclosure} and are stored in the component \texttt{allCAPs} of the output object of class \texttt{disclosure}. This is printed when the parameter \texttt{to.print} includes \texttt{allCAPs}. The first measure is known as the baseline CAP and refers to an average of the predictions that would be made by someone who only has access to the marginal distribution of the target. The intruder then guesses the CAP for each level of the target according to the relative frequencies $pd_{t.}$. Averaging this over all observations gives
\begin{equation}
  \nonumber baseCAPd = \sum{(pd_{t.})^2}.
\end{equation}
The \% of $t$ that would be correctly predicted from $q$ for someone with access to the original data is $CAPd$:
\begin{equation}
  \nonumber CAPd = \sum\limits_{tq}{(pd_{tq}d_{tq})}/N_d,
\end{equation}
and the equivalent measure for the SD treated as if it were the original is
\begin{equation}
  \nonumber CAPs = \sum\limits_{tq}{(ps_{tq}s_{tq})}/N_s.
\end{equation}
The measures  $DCAP_{b}$ and $DCAP_s$ are each based on the  percentage of  synthetic records where $t$ is correctly predicted in the SD  This gives
\begin{equation}
  \nonumber DCAP_{b} = \sum\limits_{tq}{(ps_{tq}d_{tq})}/N_{b}~~~DCAP_{s} = \sum\limits_{tq}{(ps_{tq}d_{tq})}/N_s
\end{equation}
Another option for the denominator of a $DCAP$ measure is the count of GT records, $N_d$, as used for $DiSCO$ in equation (11).
\begin{equation}
  \  DCAP_{d} = \sum\limits_{tq}{(ps_{tq}d_{tq})}/N_d
\end{equation}
When the SD and GT have the same number of records $ DCAP_{s} =  DCAP_{d}$. But when the SD has fewer records $ DCAP_{d} $ will be reduced.

The $TCAP$ and $TCAP_{b}$ measures  use the same count of disclosive records as $DiSCO$ in their numerators, but use different denominators. $TCAP_{b}$ was 
mistakenly quoted in earlier versions of this vignette as being the measure proposed in \cite{little2022}, using $N_{b}$ as its denominator:
\begin{equation}
 TCAP_{b} = 100\sum^q{\sum^t{(d_{tq} | ps_{tq} = 1)}}/N_{b}.
\end{equation}
 but the measure used in \cite{little2022, little2025} had a smaller denominator restricted to all counts in the original data, with keys corresponding to those predicted with certainty from the SD, for any target: 
\begin{equation}
TCAP = 100\sum^q{\sum^t{(d_{tq} | ps_{tq} = 1)}}/\sum^q( d_{.q} \sum^t{(ps_{tq} = 1)}),
\end{equation}
\renewcommand{\baselinestretch}{1.0}
where the summation over $t$ in the denominator indicates columns of the $tq$ tables where any target is predicted with certainty from the SD, the quantity $N_{bp}$
defined in section 3.3.2.
\end{document}